\documentclass[onecolumn,pra,superscriptaddress,floatfix,showpacs,aps,letterpaper,preprint,nobalancelastpage,accepted=2019-10-26]{quantumarticle}
\pdfoutput=1
\usepackage[numbers,sort&compress]{natbib}
\usepackage{color} 
\usepackage{transparent}
\usepackage{amsmath}
\usepackage{hyperref} 
\usepackage{amssymb}
\usepackage{amsthm}

\newtheorem{definition}{Definition}[section]

\begin{document}

\title {Randomized Benchmarking as Convolution: Fourier Analysis of Gate Dependent Errors}
\date{July 16, 2021}
\author{Seth T. Merkel}
\affiliation{HRL Laboratories, LLC 3011 Malibu Canyon Road, Malibu, CA 90265 USA}
\author{Emily J. Pritchett}
\affiliation{HRL Laboratories, LLC 3011 Malibu Canyon Road, Malibu, CA 90265 USA}
\author{Bryan H. Fong}
\affiliation{HRL Laboratories, LLC 3011 Malibu Canyon Road, Malibu, CA 90265 USA}

\begin{abstract}
We show that the Randomized Benchmarking (RB) protocol is a convolution amenable to Fourier space analysis. 
By adopting the mathematical framework of Fourier transforms of matrix-valued functions on groups established in recent work from Gowers and Hatami \cite{GH15},
we provide an alternative proof of Wallman's \cite{Wallman2018} and Proctor's \cite{Proctor17} bounds on the effect of gate-dependent noise on randomized benchmarking.    We show explicitly that as long as our faulty gate-set is close to the targeted representation of the Clifford group,  an RB sequence is described by the exponential decay of a process that has exactly  two eigenvalues close to one and the rest close to zero.  This framework also allows us to construct a gauge in which the average gate-set error is a depolarizing channel parameterized by the RB decay rates, as well as a gauge which maximizes the fidelity with respect to the ideal gate-set.  
\end{abstract}
\maketitle

\section{Introduction}
Randomized benchmarking (RB) \cite{Emerson05,Levi07,Knill08,Dankert09,Magesan11} is a workhorse of the quantum characterization community.  Used to bound errors in a variety of physical implementations of quantum processors \cite{Chow09,Ryan09,Olmschenk10,Gaebler12,Corcoles13,Barends2014,Veldhorst14,Xia15,Gaebler16,Ballance16},  RB has been expanded broadly from its original assumptions of errorless control, and depolarizing, gate-independent noise in an effort to quantify a wide variety of more-realistic error models \cite{Magesan12b,Gambetta12,Wallman15,Fogarty15,CarignanDugas15,Cross16,Wallman16,Fong17,Wood18}.  Making 
rigorous the analyses of these more-realistic models is still an active area of research \cite{Magesan11,Magesan12,Proctor17,Wallman2018}.  Of particular interest in this manuscript are RB sequences with gate-dependent errors, that is, each individual physical gate in the RB gate set is associated with its own independent error process.  

In the initial attempt to bound the effect of gate-dependent errors, Magesan {\it et al.}~use a linearization technique to treat gate-dependent errors as a perturbation with respect to a uniform error channel \cite{Magesan11,Magesan12}.   This approach defines gate error relative to a fixed representation of the operations being benchmarked, which is problematic because RB decay rates are invariant under transformations of this representation, resulting in very loose bounds on RB decay with respect to the gate error.   Chasseur and Wilhelm \cite{Chasseur15} analyzed non-perturbative gate-dependent error in the context of a modified RB protocol accounting for leakage errors.
Roughly parallel work from Wallman \cite{Wallman2018} and Proctor {\it et al.} \cite{Proctor17} give explicit examples where perturbation terms are non-negligible and where the Magesan bounds are too loose to be practically useful; they additionally justify the exponential decay of RB.  The revised methods in both manuscripts, though different in detail, involve deriving the average of RB decay sequences from what is essentially the power of a matrix. This guarantees that for generic, gate-dependent noise, the benchmarking decay will always look like the sum of two exponentials, with small corrections, independent of the gate fidelity with respect to the Clifford group in any fixed representation.

Here we develop an alternative proof that emphasizes clarity and intuition over mathematical rigor, showing that RB can be described as a convolution, and therefore some of its properties are more transparent in a Fourier space.  We use a Fourier transform from Gowers and Hatami \cite{GH15}, which extends some techniques from previous work by Moore and Russell \cite{Moore2015}.  This transform maps matrix-valued functions that act on the elements of a general group onto matrix-valued functions of the group's irreducible representations, and it has all the properties of a traditional Fourier transform -- an inverse, a convolution identity, and Parseval's theorem --- which allow us to formalize and simplify RB more naturally.  
In addition, this Fourier analysis provides the tools to construct gauge (i.e.,~similarity) transformations in which either the average gate error channel is a generalized depolarizing channel fully characterized by the RB decay rates or the average gate fidelity is maximized.  We believe that in the latter case this is the first such construction.  

The outline of this manuscript is as follows: in Section \ref{S:RBreview}, we review the basics of randomized benchmarking and show that an RB sequence can be thought of as a convolution; in Section \ref{S:Fourier}, we review matrix-valued Fourier transforms; in Section \ref{S:CliffordFourier} we apply this Fourier transformation to the super-operator representation of the Clifford group;  in Section \ref{S:RBgatedependent}, we compactly reproduce Wallman's proof of the effects of gate-dependent noise; in Section \ref{S:Eigenvecs}, we show how the eigenvectors of the Fourier transform can be used to construct gauges; finally, in Section \ref{S:Examples}, we apply this Fourier technique to reproduce examples from Proctor \cite{Proctor17} and Wallman \cite{Wallman2018}, as well as an example of our own exploring leakage characterization and the relevance of global phases to the Clifford group.         
   
\section{Randomized benchmarking as convolution}\label{S:RBreview}

In this section we review the basics of randomized benchmarking and introduce some notation.  Quantum information theorists sometimes fail to distinguish between groups and representations, but we will make their distinction explicit.  Consider the operation of a quantum processor as a function $\phi : U(2^n) \rightarrow {\mathcal Q}(2^n)$, mapping elements of the unitary group on $n$-qubits, $U(2^n)$, to the space of quantum processes, ${\mathcal Q}(2^n)$.  This mapping is consistent with Markovian error processes (otherwise we might parameterize our maps by some side-channel information, i.e.,  $\phi=\phi(u,\vec{\alpha})$) and in principle allows for leakage by the projection of a larger map to the computation subspace.  ${\mathcal Q}(2^n)$ is the space of completely-positive, trace non-increasing maps, whose elements can be expressed as ${\mathbb R}^{4^n\times 4^n}$ matrices using the standard super-operator description of a quantum processes in the computational basis (e.g., Liouville, natural, and Pauli transfer matrix representations).  In this way we can think of the operation of our quantum processor as a matrix-valued function of a group.

In any practical quantum computing application we restrict ourselves to a finite number of fundamental quantum operations, and likewise it can be useful to try to benchmark our quantum processor by its behavior with respect to a finite group.  In this manuscript we will assume we are benchmarking with respect to the Clifford group, ${\cal C}$, though the presented techniques are more general.  Randomized benchmarking consists of the following:
\begin{enumerate}
\item Prepare the system in the state $\vert \rho \rangle$.
\item Sequentially apply $m-1$ gates $\phi(C_1)$, $\phi(C_2)$, \ldots, $\phi(C_{m-1})$, sampled uniformly from the Clifford group; the associated operator product is $\phi(C_{m-1}) \ldots \phi(C_1)$.
\item Apply a final operation that ideally inverts the first $m-1$ gates, $\phi(C_1^{-1} \ldots C_{m-1}^{-1})$.
\item Make a measurement, $\langle M \vert$, that (hopefully) has some overlap with the initial state. 
\item Repeat 1--4 to obtain the survival probability, which asymptotically approaches
 \begin{equation}
 S_m =  \mathbb{E}_{C_1 \in {\cal C}} \ldots \mathbb{E}_{C_{m-1} \in {\cal C}} \langle M \vert \phi(C_1^{-1} \ldots C_{m-1}^{-1}) \phi(C_{m-1}) \ldots \phi(C_1)  \vert \rho \rangle, \nonumber
 \end{equation}
 where $\mathbb{E}_{C \in {\cal C}}$ denotes an average over the Clifford group. 
\item Repeat for different $m$ in order to fit $S_m$ to some exponential decay model.
\end{enumerate}

Ignoring preparation and measurement, we note that the expectation over quantum processes is itself a matrix-valued function of a group element $C$,
\begin{equation} 
\Phi_m(C) = \mathbb{E}_{C_1 \in {\cal C}} \ldots \mathbb{E}_{C_{m-1} \in {\cal C}}~  \phi(C C_1^{-1} \ldots C_{m-1}^{-1}) \phi(C_{m-1}) \ldots \phi(C_1),
\end{equation}
though in standard randomized benchmarking we only evaluate $C= e$ (the group identity element).  There is, however, a natural re-indexing of this expression,
\begin{equation} 
\Phi_m(C) = \mathbb{E}_{C_1 \in {\cal C}} \ldots \mathbb{E}_{C_{m-1} \in {\cal C}}~  \phi(C C_{m-1}^{-1}) \phi(C_{m-1} C_{m-2}^{-1}) \ldots \phi(C_2 C_1^{-1})\phi(C_1),
\end{equation}
that now looks like a nested series of convolutions.  In the next section, we will describe a Fourier technique that transforms matrix-valued functions of a group to matrix-valued functions of that group's irreducible representations, $\sigma$.  In this Fourier space convolutions are mapped to products, and therefore  
\begin{equation}\label{E:FtransformRB} 
\tilde{\Phi}_m(\sigma) = \tilde{\phi}(\sigma)^m,
\end{equation} 
where tilde denotes the Fourier transform.  In the limit of Markovian noise, the exponential decay of an RB sequence (i.e., the observation from Proctor and Wallman that RB is described by a matrix power) is a direct consequence of it being a convolution.  The exact form of decay depends completely on the spectrum of $ \tilde{\phi}$, the Fourier transform of our faulty gate set, which we will discuss in some detail in Sec.~\ref{S:RBgatedependent}.

\section{Fourier transforms for matrix-valued functions on finite groups}\label{S:Fourier}

Here we will briefly review Section 3 of Gowers and Hatami \cite{GH15} -- which itself is in part a review and a consolidation of notation --  covering Fourier transforms on matrix-valued functions of finite groups.  
\begin{definition}
Let ${\cal G}$ be a finite group, let $\phi : {\cal G} \rightarrow {\mathbb C}^{d_\phi \times d_\phi}$ be a matrix-valued function, and let $\sigma: {\cal G} \rightarrow U(d_\sigma)$ be an irreducible unitary representation.  The Fourier transform of $\phi$ on $\sigma$ is an $d_\phi d_\sigma \times d_\phi d_\sigma$ matrix  
\begin{equation}
\tilde{\phi}(\sigma) = {\mathbb E}_{g\in {\cal G}}~ \phi(g) \otimes \sigma^{*}(g).
\end{equation} 
\end{definition}
\noindent Gowers and Hatami show that this somewhat strange object has analogs of all the properties we would like a Fourier transform to have, namely:
\begin{enumerate}
\item(Parseval's identity 1) 
\begin{equation}\label{E:Parseval1}
 {\mathbb E}_g \| \phi(g)\|_{\rm HS}^2 = \sum_\sigma d_\sigma \| \tilde{\phi}(\sigma)\|_{\rm HS}^2 
 \end{equation}
 \item(Parseval's identity 2) 
 \begin{equation}\label{E:Parseval2}
 {\mathbb E}_g {\rm Tr}\left( \phi(g) \eta^\dagger(g) \right) = \sum_\sigma d_\sigma  {\rm Tr}\left( \tilde{\phi}(g) \tilde{\eta}^\dagger(g) \right) 
 \end{equation}
 \item(Convolution formula) 
 \begin{equation}
 \widetilde{ \phi * \eta }(\sigma)= \tilde{\phi}(\sigma) \tilde{\eta}(\sigma)  
 \end{equation}
 \item(Inverse Fourier transform) 
 \begin{equation}
 \phi(g) = \sum_\sigma d_\sigma {\rm Tr}_\sigma \left( {\mathbb I}\otimes \sigma^*(g^{-1}) \tilde{\phi}(\sigma) \right) 
 \end{equation}\label{E:U2norm}
 \item ($U_2$ norm identity)
 \begin{equation} 
 \|\phi \|^4_{U_2} = \sum_\sigma d_\sigma \| \tilde{\phi}(\sigma) \|^4_\square  
 \end{equation}
\end{enumerate}
where $\Sigma_\sigma$ denotes sums over all inequivalent irreducible representations of the group ${\cal G}$, and ${\rm Tr}_{\sigma}$ is the partial trace over the second subsystem.  We include item 5 for completeness although it's not necessary for this proof; without formally defining the $U_2$ or box norms, we just mention that they involve the sum of singular values to the fourth power.  The only norms we require in this manuscript are the Hilbert-Schmidt norm $\| \cdot \|^2_{\rm HS}$, the sum of squares of the singular values, and the operator norm $\| \cdot \|_{\rm op}$, the maximum of the singular values.

The main result of Gowers and Hatami manuscript is a stability theorem.  Broadly speaking, it states that if a function mapping a group to matrices is approximately a homomorphism, $\| \phi(g_1 g_2) - \phi(g_1)\phi(g_2) \|_{\rm HS} < \epsilon$ for every $g_1,g_2 \in {\cal G}$, then $\phi$ must be close to a (not-necessarily irreducible) representation of the group $\rho$, $\| \phi(g) - U^\dagger \rho(g) U \|_{\rm HS} < \delta$ for every $g \in {\cal G}$.  Interestingly, $\phi$ and $\rho$ may not have  the same dimension, and thus $U$ is not necessarily square.  Intuitively, we might expect an RB experiment to estimate the first expression in the stability theorem, that is, the ease with which we can invert large sequences of gates determines how well we approximate a homomorphism.  The second expression is essentially an average gate fidelity with some choice of gauge given by $U$.  The stability theorem allows us to relate these two metrics, either for finite groups such as the Clifford group or more generally for compact groups such as the special unitary group.  One minor caveat is that the stability theorem only applies if $\| \phi(g)\|_{\rm op} \leq 1$, which is not always the case for quantum processes (e.g., the amplitude damping channel), but many of the proof techniques are applicable in the following analysis.

\section{Fourier transform of the ideal Clifford group}\label{S:CliffordFourier}

Before characterizing the Fourier transform of a faulty implementation of the Clifford group,  we should understand what to expect in the ideal case.  Let's start with some useful properties of this Fourier transform when it is applied to representations themselves.  First off, {\it the Fourier transform of a representation of a group is a projector}.  To show this, assume $\phi$ is a representation of ${\cal G}$, then
\begin{align}
\tilde{\phi}(\sigma)^2 &= \widetilde{\phi * \phi}(\sigma)={\mathbb E}_{g_1} {\mathbb E}_{g_2} \phi(g_1 g_2^{-1}) \phi(g_2) \otimes \sigma^*(g_1),& \quad {\rm (convolution~identity)}\nonumber\\
&= {\mathbb E}_g \phi(g) \otimes \sigma^*(g)=\tilde{\phi}(\sigma). &\quad{\rm (definition~of~representation)}\nonumber
\end{align}
It is worth noting that the converse is not true; all projectors in Fourier space do not invert to group representations.

But what if $\phi$ is an irreducible representation?  In that case, {\it the Fourier transform  $\tilde{\phi}(\sigma)$ is a rank-1 projector $\vert \psi \rangle \langle \psi \vert$ if $\phi$ and $\sigma$ are equivalent representations, and it is zero otherwise}.  Here equivalency is defined up to a similarity transform, i.e., $\phi$ and $\sigma$ are equivalent iff $\phi = S \sigma S^{-1}$ for some $S$.  We can determine the rank of the projector through the trace and the orthogonality of characters as follows: 
\begin{align}
{\rm Tr} \left(\tilde{\phi}(\sigma)\right) &= {\rm Tr} \left( {\mathbb E}_g~ \phi(g) \otimes \sigma^*(g) \right),& \nonumber\\
&=  {\mathbb E}_g~ {\rm Tr}(\phi(g))  {\rm Tr}(\sigma^*(g)), & \quad {\rm (linearity)} \nonumber\\
&= {\mathbb E}_g \chi_\phi(g) \chi^*_\sigma(g), &\quad{\rm (definition~of~character)} \nonumber\\
&= \delta_{\phi,\sigma}. &\quad {\rm (orthonormality~of~characters~under~group~expectation)}\nonumber
\end{align}
Furthermore, we observe that  the partial trace of $\tilde{\phi}(\sigma)$ is a maximally mixed state:
\begin{equation}
{\mathbb I} = \phi(e) = \sum_{\sigma'} d_{\sigma'} {\rm Tr}_{\sigma'} \tilde{\phi}(\sigma') = d_\sigma {\rm Tr}_\sigma \tilde{\phi}(\sigma)\Rightarrow  {\rm Tr}_\sigma \tilde{\phi}(\sigma) = {{\mathbb I}}/{d_\sigma},
\end{equation}  
implying  $\vert \psi \rangle \langle \psi \vert$ must have full Schmidt rank.  In other words,  this projector $\vert \psi \rangle \langle \psi \vert$ -- the non-vanishing component of the Fourier transform of an irreducible representation --  is one very familiar to quantum information theorists, namely, it is locally equivalent to the maximally entangled bi-partite pure state $\vert \Phi \rangle = \frac{1}{\sqrt{d_\phi}} \sum_j \vert j \rangle \vert j \rangle$, but with respect to a more generic local similarity transformation as opposed to a local unitary transformation.  
 
In the super-operator representation, the Clifford group on $n$-qubits is a direct sum of two irreducible representations: the identity irrep, $\sigma_{\mathbb I}$, (i.e., the identity Pauli operator is preserved by unitary operations) and a $4^n-1$ dimensional irrep, $\sigma_{\rm P}$, (i.e., there exists some Clifford that maps every Pauli string to any other Pauli string excluding the identity).  In the ideal case our only non-zero Fourier components are both rank-1 projectors given by
\begin{align}\label{E:idealFourier}
\tilde{\phi}_{\rm ideal} (\sigma_{\mathbb I}) &=\vert \psi_{\mathbb I}  \rangle \langle \psi_{\mathbb I} \vert {\rm ~~ and ~~}
\tilde{\phi}_{\rm ideal} (\sigma_P) =  \vert  \psi_{P}\rangle \langle\psi_{P} \vert,
\end{align}     
where $\vert \psi_{\mathbb I} \rangle$ is a length $4^n$ vector of the form $1\oplus {\bf 0}_{4^n-1}$ (a one followed by $4^n-1$ zeros)and  $\vert \psi_P \rangle$ is a length $4^n(4^n-1)$ vector given by ${\bf 0}_{4^n-1} \oplus \vert \Phi \rangle$ ($4^n-1$ zeros prepended to a maximally entangled state on a $(4^n-1) \times (4^n-1)$ dimension Hilbert space).  We have included all the irreducible representations of the single qubit Clifford group and its character table in  appendix \ref{Appendix:Clifford}.   
\section{Analyzing RB with gate dependent errors}\label{S:RBgatedependent}

We can now analyze randomized benchmarking with gate dependent errors.  First, it will be useful to divide both sides of the Parseval identities (Eqs. \ref{E:Parseval1} and \ref{E:Parseval2}) by the dimension of the map $d_\phi$ (note that $d_\phi=4^n$ for an $n$ qubit system).  Rescaling the Hilbert-Schmidt norm (or trace inner product) this way defines the fidelity of entanglement, $F_e$, which is bounded above by 1 for a quantum process.  Therefore,
 \begin{equation}\label{E:quantumPars}
1 \geq  {\mathbb E}_g ~F_e \left( \phi(g), \eta(g) \right) = \sum_\sigma \frac{d_\sigma}{d_\phi}  {\rm Tr}\left( \tilde{\phi}(\sigma) \tilde{\eta}^\dagger(\sigma) \right).
 \end{equation}
Assuming that our experimental colleagues aren't just banging rocks together, $\phi$ is a decent approximation of the Clifford group, $\phi_{\rm ideal}$,  in the computational basis.  If we assume an average fidelity of $1-\delta$ we obtain,
 \begin{align}
 1-\delta &= {\mathbb E}_g ~F_e \left( \phi(g), \phi_{\rm ideal}(g) \right) =  \sum_\sigma \frac{d_\sigma}{d_\phi}  {\rm Tr}\left( \tilde{\phi}(\sigma) \tilde{\phi}_{\rm ideal}^\dagger(\sigma) \right) , & \quad {\rm \ (Eq. ~ \ref{E:quantumPars})}\nonumber\\
  &=   \frac{1}{4^n}\langle \psi_{\mathbb I} \vert  \tilde{\phi}(\sigma_{\mathbb I}) \vert \psi_{\mathbb I} \rangle +\frac{4^n-1}{4^n}\langle \psi_{P} \vert  \tilde{\phi}(\sigma_{P}) \vert \psi_{P} \rangle. & \quad {\rm \ (Eq. ~ \ref{E:idealFourier})}\label{eq:gatefid}
 \end{align}  
It is useful to denote the diagonal matrix elements $t\equiv\langle \psi_{\mathbb I} \vert  \tilde{\phi}(\sigma_{\mathbb I})\psi_{\mathbb I}\rangle$ and $p \equiv\langle \psi_{P} \vert  \tilde{\phi}(\sigma_{P}) \vert \psi_{P}\rangle$.  As a consequence of complete positivity $(p\leq t)$ and the trace non-increasing property of quantum maps $(t\leq1)$ (see Appendix \ref{Appendix:pandt}), we can bound 
\begin{equation}\label{E:bounds}
t \geq 1-\delta  \quad {\rm and} \quad p \geq 1- \delta \frac{4^n}{4^n-1} , 
\end{equation}
i.e., $p$ and $t$ are both fairly close to 1.

 The largest singular values of the Fourier matrices, $\tilde{\phi}(\sigma_{\mathbb I})$ and $\tilde{\phi}(\sigma_{P})$, are lower bounded by $t$ and $p$ respectively.  We can upper bound the size of the next largest singular value, $q$, in any of the Fourier matrices by assuming $q$ is the only other non-vanishing singular value.  Using Eq.~\ref{E:Parseval1} we have,
 \begin{equation}
1 \geq  {\mathbb E}_g ~F_e \left( \phi(g), \phi(g) \right) =  \sum_\sigma \frac{d_\sigma}{d_\phi} \| \tilde{\phi}(\sigma) \|_{\rm HS}^2 \geq  \frac{t^2 + (4^n-1)p^2 + d_{\sigma}q^2}{4^n},
\end{equation}
where the maximum $q$ for $t$ and $p$ consistent with Eq.~\ref{eq:gatefid} is given by $t=p=1-\delta$, or
\begin{equation}
q \leq \sqrt{ \frac{4^{n} (2 \delta-\delta^2)}{d_\sigma}}.
\end{equation}
While the exponential scaling in $n$ is scary and reminiscent of diamond-norm bounds on average fidelity, bounds on $q$ are actually quite reasonable for small $n$; e.g., $\delta \ll 13.3\%$ and $\delta \ll 3.1\%$ are enough to ensure that $q \ll 1$ for one- and two-qubit systems, respectively.  Tighter bounds probably exist if we restrict to Fourier transformations generated by completely positive process matrices.

At this point our proof is essentially finished, with all the heavy lifting done by Parseval's identity.  In the small error limit, our Fourier transform has a good unit-rank approximation in both the $\sigma_{\mathbb I}$ and  $\sigma_{P}$ representations.  This implies that there can at most be one eigenvalue that is not (nearly) zero in each of these irreps, and we will call these eigenvalues $\bar{t}$ and $\bar{p}$.   It would be convenient if $\bar{t}$ and $\bar{p}$ were bounded by the diagonal matrix elements $t$ and $p$ relative to some fixed choice of gauge for $\phi_{\rm ideal}$, but we will show that this is not generally true in Sec.~\ref{S:Examples}.   As we look at longer RB sequences, or raise the Fourier transforms to higher powers, our spectrum will be dominated by these two eigenvalues to ${\mathcal O}(\delta^{m/2})$.    Since both the inverse Fourier transform and the final expectation value are linear operations we find that 
\begin{equation}
S_m = A  {\bar t}^m + B {\bar p}^m +{\mathcal O}(\delta^{m/2}), 
\end{equation}
which is what we set out to show:  that randomized benchmarking generically follows an exponential decay parameterized by at most two rates.   

\section{Gauges and Eigenvectors}\label{S:Eigenvecs}

We have completed our proof using the spectrum of the Fourier transform, but before moving onto examples we should briefly discuss the related eigenvectors of the Fourier transform  and how we can use them to construct gauge transformations.  Following Gowers and Hatami \cite{GH15}, we can vectorize the Fourier transform to rewrite the Fourier eigen-equation as a matrix equation:
\begin{align}
\lambda {\bf v} =\tilde{\phi}(\sigma){\bf v} = {\mathbb E}_{g\in {\cal G}}~ \phi(g) \otimes \sigma^{*}(g) {\bf v} \iff \lambda V = {\mathbb E}_{g\in {\cal G}}~ \phi(g) V \sigma^{\dagger}(g) \label{Eq:eigen},
\end{align}
where $V$ is a $d_{\phi} \times d_{\sigma}$ matrix that contains the $d_{\phi} d_{\sigma}$ elements of the eigenvector ${\bf v}$.  We can choose $\| {\bf v} \| = d_\sigma$ as the normalization for ${\bf v}$ for reasons that will soon become apparent.  By joining the two dominant eigen-equations from the previous section we can rewrite Eq.~\ref{Eq:eigen} as
\begin{equation}
 S_{\rm dep} {\mathcal D}_{\bar{p},\bar{t}} = {\mathbb E}_{g\in {\cal G}}~ \phi(g) S_{\rm dep} \phi_{\rm ideal}^{\dagger}(g).
\end{equation} 
where we define the two $d_{\phi}  \times d_{\phi} $ matrices ${\mathcal D}_{\bar{p},\bar{t}}$ and $S_{\rm dep}$ as
 \begin{align}
{\mathcal D}_{\bar{p},\bar{t}} \equiv \left(\begin{array}{cc}
\bar{t} & {\bf 0} \\
{\bf 0} & \bar{p} {\mathbb I}_{d_{\phi}-1}
\end{array}\right) \quad {\rm and} \quad   S_{\rm dep} \equiv \left(\begin{array}{c|c}
V_{\bar{t}} & V_{\bar{p}} 
\end{array}\right).
\end{align}
The expression $\left(V_{\bar{t}} \vert V_{\bar{p}}  \right)$ denotes a matrix where the column vector $V_{\bar{t}}$  has been prepended to the columns of $V_{\bar{p}}$.  Our choice of the eigenvector normalization ensures that in the small-error limit $S_{\rm dep}$ is close to the identity, i.e., full-rank and invertible, and therefore
\begin{equation}
{\mathcal D}_{\bar{p},\bar{t}}  = {\mathbb E}_{g\in {\cal G}}~ S_{\rm dep}^{-1} \phi(g) S_{\rm dep} \phi_{\rm ideal}^{\dagger}(g) .
\end{equation} 
The eigenvectors corresponding to the eigenvalues $\bar{t}$ and $\bar{p}$ provide a unique similarity, or gauge, transformation in which the average of the individual gate error channels is a generalized depolarizing channel (i.e., a depolarizing map composed with a channel that uniformly decreases the trace) with parameters $\bar{t}$ and $\bar{p}$.  We define, $\phi_{\rm dep}(g) \equiv S_{\rm dep}^{-1} \phi(g) S_{\rm dep}$, the gate-set in the depolarizing gauge, and in this gauge the average fidelity of entanglement is given by 
\begin{equation}
 {\mathbb E}_g ~F_e \left( \phi_{\rm dep}(g), \phi_{\rm ideal}(g) \right) = \frac{\bar{t} + (4^n-1)\bar{p}}{4^n}.
\end{equation}

It is tempting to suggest that the gauge $S_{\rm dep}$ is {\it optimal} -- meaning that it maximizes the gate fidelity -- but this is not generally true.  Consider Eq.~\ref{eq:gatefid} with a general gauge transformation $S$:
\begin{equation}
{\mathbb E}_g ~F_e \left( S^{-1}\phi(g)S, \phi_{\rm ideal}(g) \right) =   \frac{1}{4^n}\langle \psi_{\mathbb I}(S) \vert  \tilde{\phi}(\sigma_{\mathbb I}) \vert \psi_{\mathbb I}(S) \rangle +\frac{4^n-1}{4^n}\langle \psi_{P}(S) \vert  \tilde{\phi}(\sigma_{P}) \vert \psi_{P}(S) \rangle\label{eq:quadform}.
\end{equation} 
Using the cyclic property of the trace we can instead apply the transformation $T^{-1}$ to the ideal gate-set, which can't change $\phi_{\rm ideal}$'s irrep decomposition, and define $|\psi_{\mathbb I}(S)\rangle$ and $|\psi_{P}(S)\rangle$ as the non-trivial eigenvectors of $\widetilde{(S\phi_{\rm ideal} S^{-1})}$.  Fourier transform matrices may not be diagonalizable, and therefore the quadratic forms in Eq.~\ref{eq:quadform} are not generally bounded by the maximum eigenvalues $\bar{t}$ and $\bar{p}$.  We can, however, construct the \emph{optimal} gauge transformation, $S_{\rm opt}$, leading to process matrices $\phi_{\rm opt}(g)$, by the observation that quadratic forms are invariant under symmetrization, and so instead of constructing a similarity transformation from the eigenvectors of $\tilde{\phi}(\sigma)$ we could instead use the eigenvectors of $( \tilde{\phi}(\sigma)+\tilde{\phi}(\sigma)^T)/2$, which is always diagonalizable.  This similarity transformation will maximize the average gate fidelity, but since the average error channel is not necessarily a generalized depolarizing channel, this reduced error rate is not easily extracted from repeated applications of the gate-set \cite{Proctor17,Wallman2018,CarignanDugas18,Qi19}.  We expect that in the small-error limit most gate-set Fourier transforms are nearly diagonalizable, that is they are nearly rank-1 with a large diagonal matrix element, and therefore $S_{\rm dep} \approx S_{\rm opt}$ to ${\mathcal O}(\sqrt{\delta})$.  Additionally, for either the depolarizing or optimal gauge transformations the transformed gate sets may no longer be completely positive. 

\section{Examples}\label{S:Examples}
In this section we will look at three examples of cases where the standard analysis of RB becomes complicated.  The first two examples are taken from the literature, both showing how fairly simple error models can lead to RB decays that are not commensurate with the average gate fidelity.  The third example describes an ideal gate-set acting on a system with a leakage level.  We treat these examples numerically, including a Mathematica notebook detailing these calculations, as well details on Clifford irreps and Fourier transforms, in the supplementary material \cite{AncillaryFile}.

\subsection{Example 1 from Proctor}

In {\it Example 1} of Proctor \cite{Proctor17}, the Clifford group is generated by composite pulse sequences of faulty $X_{\pi/2}$ and $Y_{\pi/2}$ gates.  The error in this case is a small $z$-rotation appended to each generator, i.e.~$X_{\pi/2} = \exp (-i  \theta \frac{\sigma_z}{2} ) \exp(-i \frac{\pi}{2} \frac{\sigma_x}{2} )$ and $Y_{\pi/2} = \exp (-i  \theta \frac{\sigma_z}{2} ) \exp(-i \frac{\pi}{2} \frac{\sigma_y}{2} )$.  Physically, this is a coherent memory error caused by something like a detuning or mis-timing.  There is a gate dependence in this error model because Clifford gates are not all composed from a uniform number of composite pulses. We note that we are not sure our decomposition of the Clifford gates into $X_{\pi/2}$ and $Y_{\pi/2}$ rotations is exactly the same as the decomposition in Proctor (see Appendix \ref{Appendix:Clifford}) but any differences seem to have a very small effect on the numerical outcome. 

We consider the case where $\theta=0.1$ (as in Proctor) where we find that $ {\mathbb E}_g ~F_e \left( \phi(g), \phi_{\rm ideal}(g) \right) = 1-3.70 \times 10^{-3}$.   The largest eigenvalues of our Fourier decomposition, and thus the RB decay rates are $\bar{t}=1$ and $\bar{p} =1-2.94\times 10^{-5}$, and yield to an RB estimate of ${\mathbb E}_g ~F_e \left( \phi_{\rm dep}(g), \phi_{\rm ideal}(g) \right) = 1-2.20 \times 10^{-5}$.  This two order-of-magnitude discrepancy between RB estimate and average fidelity is in agreement with the previous simulations.  The next largest eigenvalue of the Fourier transform, $\tilde{\phi}$, is $1.88\times10^{-3}$ and so we can confidently model the RB decay as a single exponential.

From this analysis we obtain the depolarizing similarity transformation 
\begin{equation}
S_{\rm dep} = \left( \begin{array}{cccc}
1 & 0 & 0 & 0 \\ 
0 & 0.997701& -0.0516457 &-0.0439113\\
0 & 0.0492626 & 0.997353 & -0.0533756\\
0 & 0.0465462 & 0.0510903 & 0.997612 \end{array}\right),
 \end{equation}
  which is nearly, but not quite, a unitary matrix.  The resulting  process matrices are nearly completely positive, but with negative Choi matrix eigenvalues of about the same order of magnitude as $1-F_e$.  The optimal gauge transformation is given by
 \begin{equation}
S_{\rm opt} = \left( \begin{array}{cccc}
1 & 0 & 0 & 0 \\ 
0 & 0.9976 & -0.0509382 & -0.0469065\\
0 & 0.0484868 & 0.997477 & -0.0518515\\
0 & 0.0494286 & 0.0494542 & 0.997552\end{array}\right),
 \end{equation}
with ${\mathbb E}_g ~F_e \left(\phi_{\rm opt}(g), \phi_{\rm ideal}(g) \right) = 1-1.62\times 10^{-5}$, only a modest improvement over the RB estimate in this case.

\subsection{Example from Wallman}

While Proctor showed an example where the average overlap with the Ideal Clifford in the computational basis overestimates the decays in RB, Wallman showed an example where the opposite can be true \cite{Wallman2018}.  Wallman's error map is that every gate is affected by a uniform depolarizing channel (a map that preserves the identity and shrinks every other Pauli element by $\nu$), and half of the Cliffords experience an additional $z$-error (again parameterized by $\theta$, but now applied to the Clifford and not the generators).  By varying which half of the Clifford's we apply the $z$-error to, we obtain a family of error channels, all of which have the same average gate error in the computational basis.

In accordance with Wallman's example, we choose $\nu=0.99$ and $\theta=0.09$ and sample 10,000 instances of the error channel out of the ${24 \choose 12}$ possible ways to apply $z$-errors to half of the Cliffords.  As expected there is no variance in the average gate error in the computational gauge, which is given by $ {\mathbb E}_g ~F_e \left( \phi(g), \phi_{\rm ideal}(g) \right) = 1-8.50 \times 10^{-3}$.  The error rate derived from RB and the depolarizing frame is very similar in the average case, ${\mathbb E}_g ~F_e \left( \phi_{\rm dep}(g), \phi_{\rm ideal}(g) \right) = 1-\left(8.50 \pm 0.12  \right) \times 10^{-3}$, but can either over- or under-estimate the error in the computational basis.  In the 10,000 trials the maximum over- and under-estimation from the computational gauge errors were less than 5\% of the total error.  We also calculate average gate error in the optimal frame and found that ${\mathbb E}_g ~F_e \left(\phi_{\rm opt}(g), \phi_{\rm ideal}(g) \right) = 1-\left(8.24 \pm 0.06 \right)\times 10^{-3}$.  The distribution of average errors in the optimal frame is somewhat tighter, but the error can still vary a significant amount.  In all cases the optimal gauge provides a lower error rate than either the computational of depolarizing gauges, as expected.

\subsection{Leakage characterization}

The final example in this manuscript doesn't explore an error process per se, but instead we examine the embedding of a qubit into a qutrit, a standard technique in characterizing leakage errors in superconducting \cite{Epstein14,Chasseur15} and semiconducting qubit implementations \cite{Ladd12}.  In the ideal case we implement this embedding as a mapping from the 24 single-qubit Clifford unitary matrices to qutrit matrices that act like the identity on the leakage space, that is $C_j \rightarrow C_j \oplus 1$.  The corresponding process matrices are now $9 \times 9$, and we can use the Gell-Mann matrices as a basis for expansion as opposed to the Pauli matrices.

Even in the case of perfect gates a peculiar thing happens: there are now many non-zero eigenvalues in the gate-set Fourier transform.  This is because the mapping described in the previous paragraph is not a representation of a group, and therefore its Fourier transform will not be a projector.   The special unitary group is a double cover, e.g.,  $X_{\pi} X_{\pi} = -{\mathbb I}$, and in the embedding we have chosen this global phase becomes a relative phase between the logical and leaked spaces and cannot be ignored.

In the qutrit embedding, the group generated by $X_{\pi/2}$ and $Y_{\pi/2}$ is the 48-element group CSU$(2,3)$ as opposed to the 24-element Clifford group $S_4$.  CSU$(2,3)$ shares all five of $S_4$'s irreducible representations but has three additional irreps that are not present in the smaller group.  One such unshared irrep we call $\sigma_u$, and is generated by the unitary representation of the Clifford gates: $X_{\pi/2} = e^{-i (\pi/2) (\sigma_x/2)}$ and $Y_{\pi/2} = e^{-i (\pi/2) (\sigma_y/2)}$.  One might think that this would necessitate the use of CSU$(2,3)$ in all cases, qubit or qutrit, but note, we never used the bare unitary representation of the group in the preceding analysis, only the process matrices.  Constructing a process matrix from the unitary representation involves a tensor product of the unitary representation with itself, in the qubit case, $\sigma_u \otimes \sigma_u = \sigma_{\mathbb I} \oplus \sigma_{P}$. $\sigma_{\mathbb I}$ and $\sigma_{P}$ are both irreps that are shared with $S_4$, and thus for an unembedded qubit we are able to substitute $S_4$ for the larger group since this representation has no dependence on the additional phase from $CSU(2,3)$.  

When we try to embed into a qutrit, our unitary representation is now $\sigma_u \oplus \sigma_{\mathbb I}$ and after converting to a process matrix we have $\left(\sigma_u \oplus \sigma_{\mathbb I}\right) \otimes \left(\sigma_u \oplus \sigma_{\mathbb I}\right) =  \sigma_{\mathbb I} \oplus \sigma_{P} \oplus \sigma_u \oplus \sigma_u \oplus \sigma_{\mathbb I}$.  This representation now has $\sigma_u$'s in the direct sum, and therefore what was a global phase can no longer be ignored.  Additionally, our process matrix is now the direct sum of five irreps, and therefore the Fourier transform will have five unit eigenvalues, instead of only two.  In a practical setting it's not clear that we really need to twirl over this larger group if the initial state and measurement of the RB process have no weight in the leakage subspace, but we have found it can greatly ease theoretical analysis.

\section{Concluding remarks} 
In this manuscript we have shown that randomized benchmarking is a convolution and therefore is more natural to explore with Fourier analysis.  In Fourier space, we directly see that RB with Markovian noise is described by powers of a fixed matrix, regardless of any gate-dependent noise.  When our processes are a good approximation of the Clifford group in the computational basis, this matrix has exactly two eigenvalues close to one while the rest are small, implying that the RB survival probability is always well described as a sum of two exponentials.  Additionally, this formalism allows us to construct gauge transformations that either a) map the average error operator to a general depolarizing channel parametrized by the RB decay rates or b) maximize the average gate fidelity with respect to the ideal Clifford gates in the computational basis.  We have applied this formalism to examples previously explored in the literature.

We have answered the question of ``what randomized benchmarking actually measures" as the error rate in a specific gauge -- that  in which the average error channel commutes with every group element, i.e., it is a generalized depolarizing channel -- and not in the gauge in which the error rate obtains a minimum.  It's not clear which of these quantities will be more important to the design and validation of fault-tolerant quantum processors where errors can be made approximately depolarizing through twirling in the error correction process, though for small errors we conjecture these two gauges are nearly equivalent because the Fourier transforms are always nearly invertible.      

In conclusion, matrix-valued Fourier transforms can greatly simplify the analysis of RB.  Even for simulation, it is more straightforward to numerically analyze the spectral properties of a handful of matrices than to approximate nested averages with Monte Carlo integration, though taking the Fourier transform for a group as large as the 2-qubit Clifford group is quite cumbersome.  We suspect that going forward, the techniques presented here will greatly ease explorations of non-Markovian and context-dependent noise's effect on randomized benchmarking.

\appendix
\section{Clifford group representations}\label{Appendix:Clifford}

In this appendix we review the representations of the single qubit Clifford group (both with and without a global phase).  In both cases the Clifford group has two generators corresponding to $\pi/2$ rotations which we will abbreviate as $x$ and $y$ for this appendix.  

\subsection{The single Clifford group, no phase}

The single qubit Clifford group modulo a global phase is better known as the group $S_4$, the symmetric group on four elements  (group $[ 24, 12 ]$ in the GAP numbering system \cite{GAP2018}.  We can divide this group into its conjugacy classes according to 
\begin{align}
c_0 &=e, \nonumber\\
c_1 &=x^2,~y^2,~y^3x^2y \nonumber\\
c_2 &=x,~y,~x^3,~y^3,~y^3xy,~y^3x^3y \nonumber\\
c_3 &=x^2y,~yx^2,~xy^2,~y^2x,~yxy,~y^3xy^3 \nonumber\\
c_4 &=xy,~yx,~x^3y^3,~y^3x^3,~xy^3,~y^3x,~x^3y,~yx^3
\end{align}
which yields the character table
\begin{align}
\begin{array}{|c||c|c|c|c|c|}
\hline
&c_0&c_1&c_2&c_3&c_4\\
\hline
\hline
\sigma_I&1&1&1&1&1\\
\hline
\sigma_p&1&1&-1&-1&1\\
\hline
\sigma_2&2&2&0&0&-1\\
\hline
\sigma_3&3&-1&-1&1&0\\
\hline
\sigma_P&3&-1&1&-1&0\\
\hline
\end{array}
\end{align}

A choice for the generators of these irreps is given by
\begin{align}
&\sigma_I(x) = 1&  \quad &\sigma_I(y) =1 \nonumber\\
&\sigma_p(x) = -1& \quad  &\sigma_p(y) =-1 \nonumber\\
&\sigma_2(x) = \left(\begin{array}{cc}
-\frac{1}{2} & \frac{\sqrt{3}}{2} \\
 \frac{\sqrt{3}}{2}&\frac{1}{2}
\end{array} \right)& \quad &\sigma_2(y) =\left(\begin{array}{cc}
-\frac{1}{2} &- \frac{\sqrt{3}}{2} \\
 -\frac{\sqrt{3}}{2}&\frac{1}{2}
\end{array} \right) \nonumber\\
&\sigma_3(x) = \left(\begin{array}{ccc}
-1 & 0 & 0 \\
0 & 0 & 1 \\
0 & -1 & 0 \\
\end{array} \right)& \quad &\sigma_3(y) =\left(\begin{array}{ccc}
0 & 0 & -1 \\
0 & -1 & 0 \\
1 & 0 & 0 \\
\end{array} \right) \nonumber\\
&\sigma_P(x) = \left(\begin{array}{ccc}
1 & 0 & 0 \\
0 & 0 & -1 \\
0 & 1 & 0 \\
\end{array} \right)& \quad &\sigma_P(y) =\left(\begin{array}{ccc}
0 & 0 & 1 \\
0 & 1 & 0 \\
-1 & 0 & 0 \\
\end{array} \right) \nonumber\\\
\end{align}

\subsection{The single Clifford group, global phase}

When we restore the global phase to the single qubit Clifford group we get the order 48 group CSU(2,3),  $2 \times 2$ conformal special unitary matrices acting on the finite field of three elements, or group $[48, 28]$ according to GAP.  The conjugacy classes are now given by

\begin{align}
c_0 &=e \nonumber\\
c_1 &=x^4 \nonumber\\
c_2 &=x^2,~y^2,~y^3x^2y,~x^6,~y^6,~y^7x^2y \nonumber\\
c_3 &=x,~y,~y^3x^3y,~x^7,~y^7,~y^7xy \nonumber\\
c_4 &=x^3,~y^3,~y^3xy,~x^5,~y^5,~y^7x^3y \nonumber\\
c_5 &=x^3y,~yx^3,~xy^3,~ y^3x,~x^5y,~y^7x^3,~y^5x,~x^7y^3\nonumber\\
c_6 &=xy,~yx,~x^3y^3,~~y^3x^3,~x^7y,~ y^7x,~x^5y^3,~y^5x^3\nonumber\\
c_7 &=x^2y,~yx^2,~yxy,~y^3xy^3,~xy^2,~y^2x,~x^6y,~y^5x^2,x^5y^2,~y^6x,~y^5xy,~y^7xy^3
\end{align}
which yields a character table, 
\begin{align}
\begin{array}{|c||c|c|c|c|c|c|c|c|}
\hline
&c_0&c_1&c_2&c_3&c_4&c_5&c_6&c_7\\
\hline
\hline
\sigma_I&1&1&1&1&1&1&1&1\\
\hline
\sigma_p&1&1&1&-1&-1&1&1&-1\\
\hline
\sigma_2&2&2&2&0&0&-1&-1&0\\
\hline
\sigma_u&2&-2&0&\sqrt{2}&-\sqrt{2}&-1&1&0\\
\hline
\sigma_n&2&-2&0&-\sqrt{2}&\sqrt{2}&-1&1&0\\
\hline
\sigma_3&3&3&-1&-1&-1&0&0&1\\
\hline
\sigma_P&3&3&-1&1&-1&0&0&-1\\
\hline
\sigma_4&4&-4&0&0&0&1&-1&0\\
\hline
\end{array}
\end{align}

We can generate with irreps of CSU(2,3) with exactly the same generators as $S_4$ as well as the three additional irrep generators given below (which now contain the more familiar definitions of $X_{\pi/2}$ and $Y_{\pi/2}$ in the computational basis).

\begin{align}
&\sigma_u(x) = \left(\begin{array}{cc}
\frac{1}{\sqrt{2}} &- \frac{i}{\sqrt{2}} \\
 - \frac{i}{\sqrt{2}}&\frac{1}{\sqrt{2}} 
\end{array} \right)& \quad &\sigma_u(y) =\left(\begin{array}{cc}
\frac{1}{\sqrt{2}} &- \frac{1}{\sqrt{2}} \\
  \frac{1}{\sqrt{2}}&\frac{1}{\sqrt{2}} 
\end{array} \right) \nonumber\\
&\sigma_n(x) = \left(\begin{array}{cc}
-\frac{1}{\sqrt{2}} &\frac{i}{\sqrt{2}} \\
 \frac{i}{\sqrt{2}}&-\frac{1}{\sqrt{2}} 
\end{array} \right)& \quad &\sigma_n(y) =\left(\begin{array}{cc}
-\frac{1}{\sqrt{2}} & \frac{1}{\sqrt{2}} \\
 -\frac{1}{\sqrt{2}}&-\frac{1}{\sqrt{2}} 
\end{array} \right) \nonumber\\
&\sigma_4(x) =\frac{1}{2 \sqrt{2}} \left(\begin{array}{cccc}
-1 & \sqrt{3} & i &-i\sqrt{3} \\
\sqrt{3} & 1 & -i \sqrt{3} &-i \\
 i & -i\sqrt{3} & -1 &\sqrt{3}\\
 -i \sqrt{3} & -i & \sqrt{3} &1 \\
\end{array} \right)& \quad &\sigma_4(y) =\frac{1}{2 \sqrt{2}} \left(\begin{array}{cccc}
-1 & -\sqrt{3} & 1 &\sqrt{3} \\
-\sqrt{3} & 1 &  \sqrt{3} &-1  \\
-1 & -\sqrt{3} & -1 &-\sqrt{3} \\
-\sqrt{3} & 1 &  -\sqrt{3} &1  \\
\end{array} \right) \nonumber\\
\end{align}

\section{Showing $p \leq t \leq 1$}\label{Appendix:pandt}

To show how $p \leq t \leq 1$ is implied by $\phi$ completely positivity and trace non-increasing, it helps to be more explicit in our construction of process matrices.  We define a process by its action on products of Pauli matrices, $P_j$, a complete basis for Hermitian operators.  By vectorizing or column-stacking the Pauli product matrices, $\vert P_j \rangle$, can write our quantum processes as real, $4^n \times 4^n$ matrices.  We know that $\phi_{\rm ideal}$ is composed of Clifford operators, which are unitary operations that map Pauli strings to other Pauli strings.  This implies that the process matrices for $\Phi_{\rm ideal}$ will have exactly one non-zero entry of $\pm 1$ in each row and column.  Furthermore $\phi_{\rm ideal}$ has the block structure $\sigma_{\mathbb I} \oplus \sigma_P$, with $\sigma_{\mathbb I}$ spanned by $\vert I \rangle$ and $\sigma_P$ by the remaining $4^n-1$ basis elements $\vert P_j \neq I \rangle$.  

Let's define two orthogonal projectors $\Pi_{\mathbb I} = \vert I \rangle \langle I \vert$ and $\Pi_{P} = \mathbb{I} - \Pi_{\mathbb I}$.  One can show that 
\begin{equation}
t = \langle \psi_{\mathbb I} \vert \tilde{\phi}(\sigma_{\mathbb I}) \vert \psi_{\mathbb I} \rangle = {\mathbb E}_g ~{\rm Tr} \left( \phi(g) \Pi_{\mathbb I} \phi_{\rm ideal}^\dagger(g) \Pi_{\mathbb I} \right), 
\end{equation} 
and
\begin{equation}
p= \langle \psi_{P} \vert \tilde{\phi}(\sigma_{P}) \vert \psi_{P} \rangle = {\mathbb E}_g ~{\rm Tr} \left( \phi(g) \Pi_{P} \phi_{\rm ideal}^\dagger(g) \Pi_{P} \right) / (4^n-1).
\end{equation}
Furthermore, since $\phi_{\rm ideal}$ is a unitary map, $\Pi_{\mathbb I} \phi_{\rm ideal}(g) \Pi_{\mathbb I}= \Pi_{\mathbb I}$, and we can simplify the expression for $t$ to
\begin{equation}
t  = {\mathbb E}_g ~\langle I \vert \phi(g) \vert I \rangle.
\end{equation}
If $t>1$, then there must exist a $g$ such that $\langle I \vert \phi(g) \vert I \rangle > 1$ which would violate our assumption that $\phi$ is trace non-increasing.

Showing that  $p \leq t$ involves a similar but more involved argument.  $p$ is an equally weighted average of terms of the form $\pm \langle P_j \vert \phi(g) \vert P_k \rangle$ and so it must be the case that for some $h \in {\cal G}$ and for some $P_n$ and $P_m$ not equal to the identity there must exist $\langle P_n\vert \phi(h) \vert P_m \rangle \geq p$.  We can now apply the map $\phi(h)$ to one of the two positive semidefinite operators $\vert I \rangle \pm \vert P_m \rangle$ which yields
\begin{equation}
\vert \rho \rangle = \phi(h) \left(\vert I \rangle \pm \vert P_m \rangle \right)  = t \vert I \rangle  + \sum_{P_j \neq I} c_j \vert P_j\rangle 
\end{equation}
where $c_j \equiv \langle P_j\vert \phi(h) \vert I \rangle \pm \langle P_j\vert \phi(h) \vert P_m \rangle$, and we have used the observation that a trace non-increasing map must have $\langle I \vert \phi(h) \vert P_m \rangle = 0$ (otherwise the trace of one of $\pm \vert  P_m \rangle$ would increase under the action of $\phi(h)$).  We introduced the sign ambiguity earlier so that we can ensure that $c_n = \langle P_n\vert \phi(h) \vert I \rangle \pm \langle P_n\vert \phi(h) \vert P_m \rangle$ has magnitude $|c_n| \geq p$.

To complete the argument we need to show that this $\rho$ necessarily has a negative eigenvalue which, since $\rho$ is Hermitian, can be shown by providing a $\vert \psi \rangle$ such that $\langle \psi \vert \rho \vert \psi \rangle < 0$.  Let's construct a set of $2^n -1$ commuting Pauli strings that contains $P_n$ by considering products of $P_n$'s constituent single qubit Pauli operators (replacing any identities with $Z$ operations).  As an example, if $P_n = ZIX$, we would say $P(\vec{\alpha}) = Z^{\alpha_0} Z^{\alpha_1} X^{\alpha_2}$, where we've indexed these $2^n$ Pauli strings by a binary vector, $\vec{\alpha} \in \{0,1\}^{\otimes n}$.  There is a natural tensor-product basis for this set of operators, the eigen-basis of the constituent single qubit Pauli operators, which we index by another binary vector $\vec{\beta}$, e.g., $\vert \psi(\vec{\beta}) \rangle= \vert (\beta_0)_Z \rangle \vert (\beta_1)_Z \rangle \vert (\beta_2)_X \rangle$.  We can now write the expectation value,
\begin{equation}
\langle \psi(\vec{\beta}) \vert \rho \vert \psi(\vec{\beta}) \rangle = t + \sum_{\vec{\alpha} \neq 0} (-1)^{\vec{\alpha}\cdot \vec{\beta}} c_{\vec{\alpha}} ,
\end{equation}   
where we have utilized that any Pauli string outside of our commuting set has an expectation value of zero with respect to any $\vert \psi(\vec{\beta}) \rangle$.  The $c_{\vec{\alpha}}$'s may all have arbitrary signs, but it should be clear that we can choose a $\vec{\beta}$ such that all terms in the sum are negative.  That $\vec{\beta}$ will lead to a minimum,
\begin{equation}
\min_{\vec{\beta}} \langle \psi(\vec{\beta}) \vert \rho \vert \psi(\vec{\beta}) \rangle = t - \sum_{\vec{\alpha} \neq 0} |c_{\vec{\alpha}}|   \leq  t - p,
\end{equation}
and so, if $t<p$ we are guaranteed a process matrix that does not map positive operators to other positive operators. 

\bibliographystyle{plainnat}
\bibliography{RBConvolution}

\end{document}